\documentclass[amsmath,amssymb,numbers]{revtex4}

\begin{document}

\title{Radon Transform in finite Hilbert Space}

\author{M. Revzen}
\affiliation {Department of Physics, Technion - Israel Institute of Technology, Haifa
32000, Israel}

\date{\today}

\begin{abstract}

Novel analysis of finite dimensional Hilbert space is outlined. The approach bypasses general, inherent,
difficulties present in handling angular
variables in finite dimensional problems: The finite dimensional, d, Hilbert space
operators are underpinned with finite geometry which provide intuitive perspective to the
physical operators. The analysis emphasizes a central role for projectors of mutual
unbiased bases (MUB) states, extending thereby their use in  finite dimensional
quantum mechanics studies. Interrelation among the Hilbert space operators revealed via
their (finite) dual affine plane geometry (DAPG) underpinning are displayed and utilized in
formulating the finite dimensional ubiquitous Radon transformation and its inverse
illustrating phase space-like physics encoded in lines and points of the geometry.
The finite geometry required for our study is outlined.

\end{abstract}


\maketitle

\section{Introduction}

The mathematics of Radon transform was introduced for studies of Astronomical data
\cite{radon,deans,mello3}. Today it finds an extensive application in CAT (computer
assisted tomography) scan \cite{guy,freeman,mello3}, state reconstruction
\cite{ulf,schleich,vogel,walls} as well as within the fundamental phase space formulation
of quantum mechanics \cite{kim,schleich}. The methodology is geometrically based:
A phase space image of an arbitrary Hilbert space operator $\hat{A}$  is given by \cite{moyal},

\begin{equation}
\hat{A}\rightarrow W_{A}(q,p)=\int dye^{ipy}\langle q-y/2|\hat{A}|q+y/2\rangle.
\end{equation}

The Wigner function \cite{wigner}, $W_{\rho}(q,p)$, being the image of the density
operator, $\hat{\rho}$, although {\it not} non-negative, does enjoy many of the attributes
of a phase space distribution. In particular its marginal,
\begin{equation}
\tilde{\rho} (x',\theta)=\int{ dq dp \delta(Cq+Sp-x')W_{\rho}(q,p)},\;\;  C=cos\theta, S=sin\theta,
\end{equation}
is its Radon transform \cite{schleich,yves}:
\begin{equation}\label{conrad}
{\cal{R}}[W_{\rho}](x,\theta)=\tilde{\rho}(x,\theta);\;\;{\cal{R}}^{-1}[\tilde{\rho}](q,p)=W_{\rho}(q,p).
\end{equation}

The explicit expression for $W_{\rho}(q,p)$ in terms of principal value integral of
$\tilde{\rho}(x,\theta$ is given in \cite{schleich,ulf,vogel,walls,mello3}. We offer in the
last section an interpretation for this singular relation gained via the finite dimensional
approach.\\

The tomographic luster of the Radon transform suggests viewing it via mutual unbiased bases
(MUB). Informationally complete set of MUB, i.e. such that allows complete accounting of an
arbitrary state, are $\{|x',\theta \rangle\};\;-\infty\le x'\le \infty,;\;0 \le \theta \le
\pi$. These are the eigenstates of the dynamical variables
$\hat{X}_{\theta}=C\hat{x}+S\hat{p},$ \cite{mello3,revzen1}. Now
\begin{equation}
tr\hat{\rho} \\mathbb{P}(x',\theta)=\int{ dq dp\delta(Cq+Sp-x')W_{\rho}(q,p})=\tilde{\rho}
(x',\theta),
\end{equation}
illustrating a quasi distributional attribute of $W_{\rho}(q,p)$, as the phase space
mappings of $\mathbb{P}(x',\theta)\equiv|x',\theta \rangle\langle \theta,x'|$ is
$\delta(x'-Cq-Sp).$ Noteworthy is the pivotal role of the delta function: it identifies the
Radon transformation, defines the marginal distribution and constitute
the phase space mapping of an MUB projector.\\
The present intuitive approach involves the  replacement of phase space with finite
geometry's points and lines. The role of the Dirac delta function is played by what we
signify with $\Lambda_{\alpha,j}$ where $\alpha$ and j refer to
 finite geometry point and line respectively. In this way we eschew in the finite dimensional
 intrinsically sticky issue of angular variables. On the physical side the salient feature is the
 mapping of finite dimensional Hilbert space operators onto c number functions of the phase space like
 geometrical points and lines and, in particular, the definition of quasi distribution and
 Radon transformation over this phase space like points and lines. These ideas are clarified below.\\

 \section{ Finite Geometry and Hilbert Space Operators}

We now briefly review the essential features of finite geometry required for our study
\cite{bennett,grassl,diniz,shirakova,tomer,wootters4}.\\
A finite plane geometry is a system possessing a finite number of points and lines. There
are two kinds of finite plane geometry: affine and projective. We shall confine ourselves
to affine plane geometry (APG) which is defined as follows. An APG is a non empty set
whose
elements are called points. These are grouped in subsets called lines subject to:\\
1. Given any two distinct points there is exactly one line containing both.\\
2. Given a line L and a point S not in L ($S \ni L$), there exists exactly one line L'
containing S
such that $L \bigcap L'=\varnothing$. This is the parallel postulate.\\
3. There are 3 points that are not collinear.\\
It can be shown \cite{bennett,diniz,shirakova} that for $d=p^m$ (a power of prime) APG can be
constructed (our study here is for d=p) and the following properties are, necessarily,
built in: \\
a. The number of points is $d^2;$ $S_{\alpha},\;{\alpha = 1,2,...d^2}$ and the number of
lines is d(d+1); $L_j,\;j=1,2....d(d+1)$.\\
b. A pair of lines may have at most one point in common.\\

c. Each line is made of d points and each point is common to d+1 lines:
$L_j=\bigcup_{\alpha}^d S_{\alpha}^j$, $S_{\alpha}=\bigcap_{j=1}^{d+1}L_j^{\alpha}.$\\
d. If a line $L_j$ is parallel to the distinct lines $L_k\;and\;L_i$ then $L_k \parallel
L_i$. The $d^2$ points are grouped in sets of d parallel lines. There are d+1 such
groupings.\\
e. Each line in a set of parallel lines intersect each line of any other set.\\

The existence of APG implies \cite{bennett,diniz,grassl,shirakova}the existence of its dual
geometry DAPG wherein the points and lines are interchanged. Since we shall study
extensively this, DAPG, we list the
corresponding properties for it. We shall refer to these by DAPG(y):\\
a. The number of lines is $d^2$, $L_j,\;j=1,2....d^2.$ The number of points is d(d+1),
$S_{\alpha},\;{\alpha = 1,2,...d(d+1)}.$\\
b. A pair of points on a line determine a line uniquely. Two (distinct) lines share one and only
one point.\\
c. Each point is common to d lines. Each line contain d+1 points.\\
d. The d(d+1) points may be grouped in sets, $R_{\alpha}$, of d points each no two of a set
share a line. Such a set is designated by $\alpha' \in \{\alpha \cup M_{\alpha}\},\;
\alpha'=1,2,...d$. ($M_{\alpha}$ contain all the points not connected to $\alpha$ - they
are not connected among themselves.) i.e. such a set contain d disjointed (among themselves)
points. There are d+1 such sets:
\begin{equation}
\bigcup_{\alpha=1}^{d(d+1)}S_{\alpha}=\bigcup_{\alpha=1}^d R_{\alpha};\;\;\;\;
R_{\alpha}=\bigcup_{\alpha'\epsilon\alpha\cup M_{\alpha}}S_{\alpha'};\;\;\;\;
R_{\alpha}\bigcap R_{\alpha'}=\varnothing,\;\alpha\ne\alpha'.
\end{equation}
e. Each point of a set of disjoint points is connected to every other point not in its
set.\\

DAPG(c) allows the identification, which we adopt,
\begin{equation}\label{A}
S_{\alpha}\equiv\frac{1}{d}\sum_{j\in\alpha}^{d} L_j.\;\;\Rightarrow\;\;\sum_{\alpha' \in
\alpha \cup M_{\alpha}}^{d}S_{\alpha'}=\frac{1}{d}\sum^{d^2} L_j.
\end{equation}
We list now some direct DAPG implied interrelation subjected to Eq. (\ref{A}): DAPG(d)
implies

$$\sum_{\alpha}^{d+1} \sum_{\alpha' \in \alpha \cup
M_{\alpha}}^{d}S_{\alpha'}=\sum^{d(d+1)}S_{\alpha}
                          =\frac{d+1}{d}\sum^{d^2} L_j\;\;\Rightarrow$$

\begin{equation}\label{I}
\sum_{\alpha' \in \alpha \cup M_{\alpha}}^{d}S_{\alpha'}=\frac{1}{d}\sum^{d^2} L_j=
\frac{1}{d+1}\sum^{d(d+1)}S_{\alpha}.
\end{equation}

\section{   Finite dimensional Mutual Unbiased Bases, MUB, Brief Review}

In a finite, d-dimensional, Hilbert space two complete, orthonormal vectorial bases, ${\cal
B}_1,\;{\cal B}_2$,
 are said to be MUB if and only if (${\cal B}_1\ne {\cal B}_2)$
 \cite{wootters1,wootters2,wootters3,wootters4,tal,vourdas,klimov2,bengtsson,planat1,amir,mello1,ent,peres}:

\begin{equation}
\forall |u\rangle,\;|v \rangle\; \epsilon \;{\cal B}_1,\;{\cal B}_2 \;resp.,\;\;|\langle
u|v\rangle|=1/\sqrt{d}.
\end{equation}
The physical meaning of this is that knowledge that a system is in a particular state in
one basis implies complete ignorance of its state in the other basis.\\
Ivanovic \cite{ivanovich} proved that there are at most d+1 MUB, pairwise, in a
d-dimensional Hilbert space and gave an explicit formulae for the d+1 bases in the case of
d=p (prime number). Wootters and Fields \cite{wootters2} constructed such d+1 bases for
$d=p^m$ with m an integer. Variety of methods for construction of the d+1 bases for $d=p^m$
are now available
\cite{tal,klimov2,vourdas}. Our present study is confined to $d=p\;\ne2$.\\
 We now give explicitly the MUB states in conjunction with the algebraically complete
 operators \cite{schwinger,amir} set:
 $\hat{Z},\hat{X}$.  Thus we label the d distinct states spanning the Hilbert space,
 termed
 the computational basis, by $|n\rangle,\;\;n=0,1,..d-1; |n+d\rangle=|n\rangle$
\begin{equation}
\hat{Z}|n\rangle=\omega^{n}|n\rangle;\;\hat{X}|n\rangle=|n+1\rangle,\;\omega=e^{i2\pi/d}.
\end{equation}
The d states in each of the d+1 MUB bases \cite{tal,amir}are the states of the
computational basis (CB) and
\begin{equation} \label{mxel}
|m;b\rangle=\frac{1}{\sqrt
d}\sum_0^{d-1}\omega^{\frac{b}{2}n(n-1)-nm}|n\rangle;\;\;b,m=0,1,..d-1.
\end{equation}
Here b labels the other d bases   and  m labels the states within a basis. We have
\cite{tal}
\begin{equation}\label{tal1}
\hat{X}\hat{Z}^b|m;b\rangle=\omega^m|m;b\rangle.
\end{equation}
For later reference we shall refer to the computational basis (CB) by b=-1. Thus the above
gives d+1 bases, b=-1,0,1,...d-1. The total number of states d(d+1) are grouped in d+1 sets
each of d states. We have of course,
\begin{equation}\label{mub}
\langle m;b|m';b\rangle=\delta_{m,m'};\;\;|\langle m;b|m';b'\rangle|=\frac{1}{\sqrt d},
\;\;b\ne b'.
\end{equation}
We remark at this junction that the eigen values of the CB might be considered finite
dimensional modulated position values ("q") and the eigenvalues of shifting operator, X,
modulated momentum ("p").\\
This completes our discussion of MUB.\\

\section{ DAPG underpinning of d-dimensional Hilbert space}

 Now the underpinning of Hilbert space operators with DAPG  will be undertaken.
We consider d=p, a prime, $\ne2$. For d=p we may construct d+1 MUB
\cite{ivanovich,tal,wootters1,wootters4}. Points will be associated with MUB state
projectors. To this end we recall that we designate the MUB states by $|m,b\rangle.$ with
$b=0,1,2...d-1$ labels the eigenfunction of, resp. $XZ^b$. m labels the state within a
basis. We designate the computational basis, CB, by  b=-1.( Note that the first column
label of -1 is for convenience and does not designate negative value of a number.) The
projection operator defined by,
\begin{equation}\label{ptop}
\hat{A}_{\alpha}\equiv |m,b\rangle \langle
b,m|;\;\alpha=\{b,m\};\;\;b=-1,0,1,2...d-1;\;m=0,1,2,..d-1.
\end{equation}
The point label, $\alpha=(m,b)$ is now associated with the projection operator,
$\hat{A}_{\alpha}$ . We now consider a realization, possible for d=p, a prime , of a d
dimensional DAPG,  as points marked on a rectangular whose horizontal width (x-axis) is
made of d+1 columns of points. Each column is labelled by b, and its vertical height (y
axis) is made of d points each marked with m. The total number of points is d(d+1) - there
are d points in each of the d+1 columns. We associate the d points $m=0,1,2,...d-1$ in each
set, labelled by b, ($\alpha \sim (m,b)$) to the {\it disjointed} points of DAPG(d),
$R_{\alpha}$ viz. for fixed b $\alpha' \in \alpha \cup M_{\alpha}$ form a column. The
columns are arranged according to their basis label, b. The first, i.e. left most, being
b=-1, $\alpha_{-1}=(m,-1);m=0,1,...d-1.$ Next we place the columns with increasing values
of b, the basis label, as we move to the right. Thus the right most column is for b=d-1.
Lines are now made of d+1 points, each of different b (necessarily , DAPG(b)). A point
$S_{\alpha}$ underpins a Hilbert space state projector, $\hat{A}_{\alpha}$. i.e.
$\hat{A}_{\alpha}^{2}=\hat{A}_{\alpha},$ and $tr\hat{A}_{\alpha}=1.$ We designate the line
operator underpinned with $L_j,$ by $\hat{P}_j$. Thus the above relations now hold with
$S_{\alpha}\leftrightarrow \hat{A}_{\alpha};\;\;L_j\leftrightarrow \hat{P}_j$.\\

e.g. for d=3 the underpinning's schematics is,
\[ \left( \begin{array}{ccccc}
m\backslash b&-1&0&1&2 \\
0&A_{(0,-1)}&A_{(0,0)}&A_{(0,1)}&A_{(0,2)}\\
1&A_{(1,-1)}&A_{(1,0)}&A_{(1,1)}&A_{(1,2)}\\
2&A_{(2,-1)}&A_{(2,0)}&A_{(2,1)}&A_{(2,2)}\end{array} \right)\].\\

Now DAPG(c) (and Eq.(\ref{ptop}),({\ref{I}))implies that
$A_{\alpha};\;\alpha=0,1,2...d-1;\;\;\alpha \in \alpha' \cup M_{\alpha'}$ forms an
orthonormal basis for the d-dimensional Hilbert space:
\begin{equation}
\sum_m^d |m,b\rangle\langle b,m|=\sum_{\alpha' \in \alpha\cup M_{\alpha}}^d\hat{A}_{\alpha'}
=\hat{I};\;\;\;
\sum_{\alpha}^{d(d+1)}\hat{A}_{\alpha}=(d+1)\hat{I}.
\end{equation}

Eq.(\ref{A}) implies,
\begin{equation}\label{A1}
A_{\alpha}=\frac{1}{d}\sum_{j\in\alpha}^d P_j.
\end{equation}

Evaluating
\begin{equation}\label{line}
\sum_{\alpha\in j}A_{\alpha}=\frac{1}{d}\sum_{\alpha\in j}\sum_{j'\in\alpha}P_j=
\frac{1}{d}\big[\sum^{(d-1)(d+1)}_{j'\ne j}
P_{j'}+(d+1)P_j\big]=I+P_j\;\Rightarrow\;P_j=\sum^{d+1}_{\alpha\in j}A_{\alpha}-I.
\end{equation}

 Eq.(\ref{mub}) implies,
 \begin{equation}
 tr A_{\alpha}A_{\alpha'}=\begin{cases}1;\;\;\alpha=\alpha'\\
 0;\;\;\alpha \ne \alpha';\alpha \in \alpha'\cup M_{\alpha'}.\\
 \frac{1}{d};\;\alpha\ne \alpha';\;\alpha \notin \alpha'\cup M_{\alpha'}.\end{cases}
 \end{equation}

Hence, using Eq.(\ref{A}),(\ref{mub}),
\begin{equation}\label{delta1}
tr A_{\alpha}P_j=\begin{cases}\sum_{\alpha' \ne \alpha}^{d}tr A_{\alpha}A_{\alpha'}=1;\;\alpha\in j.\\ \sum_{\alpha'\ne \alpha}^{d}tr A_{\alpha}A_{\alpha'}-A_{\alpha}=0;\;\alpha\ni j. \end{cases}
\end{equation}
Trivially
\begin{equation}
trP_j=\sum^{d+1}trA_{\alpha}-1=1.
\end{equation}
\begin{equation}
trP_j=\sum^{d+1}trA_{\alpha}-1=1.\;\;\Rightarrow\;\;trP_jP_{j'}=\sum_{\alpha'\in
j'}trP_jA_{\alpha'}-1=\begin{cases}d\;\;j=j'\\0\;\;j\ne j',\end{cases}
\end{equation}
i.e.
\begin{equation}
trP_jP_{j'}=d\delta_{j,j'}.
\end{equation}

An alternative view of the Lambda function is gained via

\begin{equation}\label{delta2}
tr A_{\alpha}P_j=\frac{1}{d}\sum^{d}P_{j'}P_j=\begin{cases}\frac{1}{d}\big(tr P_j^2+tr \sum_{j'\notin\alpha} P_{j'}P_j\big)=1;\;\;j\in \alpha \\
\sum_{j'\ne j}P_{j'}P_j=0;\;\;j\notin \alpha.\end{cases}
\end{equation}
Note that the case of $j\notin\alpha$ {\it implies} $j\in M_{\alpha}$.\\
These are summarized by

\begin{equation}\label{del1}
\Lambda_{\alpha,j}\equiv tr A_{\alpha}P_j=\begin{cases}1;\;\alpha\in j,\\
0;\;\alpha\notin j, \end{cases}=\begin{cases}1;\;\;j\in \alpha \\
0;\;\;j\notin \alpha.\end{cases}
\end{equation}

\section{  Geometric Underpinning of MUB Quantum Operators: The line operator}

We now consider a particular realization of DAPG of dimensionality $d=p,\ne 2$ which is the
basis of our present study.  Thus $\alpha=m(b)$ designate a point by its row, m, and its
column, b; when b is allowed to vary - it designate the point's row position in every
column. We now assert that the d+1 points, $m_j(b), b=0,1,2,...d-1,$  and  $m_j(-1)$, form
the line j which contain the two (specific) points m(-1) and m(0). The line is given by (we
forfeit the subscript j - it is implicit),
\begin{eqnarray}\label{m(b)}
m(b)&=&\frac{b}{2}(c-1)+m(0),\;mod[d]\;\;b\ne -1, \nonumber \\
m(-1)&=&c/2.
\end{eqnarray}

The rationale for this particular form will be clarified below. Thus a line j is
parameterized fully by $j=(m(-1),m(0))$. (Note: since our line labelling is based on b
 values -1 and 0  a more economic label for j is
$j=(m_{-1},m_0)$ i.e. the m values for b=-1 and 0. We shall use either when no confusion
should arise.) We now prove that the set $j=1,2,3...d^2$ lines
covered by Eq.(\ref{m(b)}) form a DAPG.\\
\noindent 1. Since each of the  parameters, m(-1) and m(0), can have d values the number of
lines is $d^2$; the number of points in a line is evidently d+1: a point for each b.  DAPG(a).\\
\noindent 2. The linearity of the equation precludes having two points with a common value
of b on the same line, DAPG(d). Now consider two points on a given line, $m(b_1),m(b_2);\;b_1\ne
b_2$. We have from Eq.(\ref{m(b)}), ($b\ne -1,\;b_1 \ne b_2$)
\begin{eqnarray}\label{twopoints}
m(b_1)&=&\frac{b_1}{2}(c-1)+m(0),\;\;mod[d]\nonumber\\
m(b_2)&=&\frac{b_2}{2}(c-1)+m(0),\;\;mod[d].
\end{eqnarray}
These two equation determine uniquely ({\it for d=p, prime}) m(-1) and m(0). DAPG(b).\\
\noindent For fixed point, m(b), $c\Leftrightarrow m(0)$ i.e the number of free parameters
is d (the number of points on a fixed column). Thus each point is common to d lines. That
the line contain d+1 is obvious. DAPG(c).\\
\noindent 3. As is argued in 2 above no line contain two points in the same column (i.e.
with equal b). Thus the d points, $\alpha,$ in a column form a set
$R_{\alpha}=\bigcup_{\alpha'\epsilon\alpha\cup M_{\alpha}}S_{\alpha'},$ with trivially
$R_{\alpha}\bigcap R_{\alpha'}=\varnothing,\;\alpha\ne\alpha',$ and
$\bigcup_{\alpha=1}^{d(d+1)}S_{\alpha}=\bigcup_{\alpha=1}^d R_{\alpha}.$ DAPG(d).\\
\noindent 4. Consider two arbitrary points {\it not} in the same set, $R_{\alpha}$ defined
above: $m(b_1),\;m(b_2)\;\;(b_1\ne b_2).$ The argument of 2 above states that, {\it for
d=p}, there is a unique solution for the two parameters that specify the line containing
these points. DAPG(e).\\
We illustrate the above for d=3, where we explicitly specify the points contained in the
line $j=\big(m(-1)=(1,-1),m(0)=(2,0)\big)$
\[ \left( \begin{array}{ccccc}
m\backslash b&-1&0&1&2 \\
0&\cdot&\cdot&\cdot&(0,2)\\
1&(1,-1)&\cdot&(1,1)&\cdot\\
2&\cdot&(2,0)&\cdot&\cdot\end{array} \right)\].\\
For example the point m(1) is gotten from
$$ m(1)= \frac{1}{2}(2-1)+2=1\;\;mod[3]\;\;\rightarrow\;m(1)=(1,1).$$
Similar calculation gives the other point: m(2)=(0,2). i.e. the line j=(1,2) contains the points
(1,-1),(2,0),(1,1) and (0,2).

The geometrical line, $L_j, \;j=(1,2)$ given above upon being transcribed
to its operator formula is via Eq.(\ref{line}),

\begin{equation}\label{lineop}
P_{j=(1,2)}=A_{(1,-1)}+A_{(2,0)}+A_{(1,1)}+A_{(0,2)}-\hat{I}.
\end{equation}

Evaluating the point operators, $\hat{A}_{\alpha}$,
\begin{equation}\label{point1}
A_{(1,-1)}=\begin{pmatrix}0&0&0\\0&1&0\\0&0&0\end{pmatrix},A_{(2,0)}=\frac{1}{3}\begin{pmatrix}1&\omega^2&\omega\\
\omega&1&\omega^2\\\omega^2&\omega&1\end{pmatrix},A_{(1,1)}=\frac{1}{3}\begin{pmatrix}1&\omega&\omega\\\omega^2&1&1\\
\omega^2&1&1\end{pmatrix},A_{(0,2)}=\frac{1}{3}\begin{pmatrix}1&1&\omega\\1&1&\omega\\\omega^2&\omega^2&1\end{pmatrix},
\end{equation}
and evaluating the sum, Eq.(\ref{lineop}), gives
\begin{equation} \label{pj}
P_{j:(m(-1)=1,m(0)=2)}=\begin{pmatrix}0&0&\omega\\ 0&1&0\\ \omega^2&0&0\end{pmatrix}.
\end{equation}

This operator obeys $P_{(1,2)}^2=\hat{I}$. That this is quite general, viz $P_j^2=\hat{I},\;\forall j$ is shown in \cite{revzen2}. In appendix A we show that $\hat{P}_j^2=\hat{I}\;\forall j $ implies the operator relation,
$$\sum_{\alpha \ne\alpha' \in
j}\hat{A_\alpha}\hat{A_\alpha'}=\sum_{\alpha \in j}\hat{A_\alpha}.$$ It is, perhaps,  of
interest that, if we associate the CB states with the position variable, q, of the
continuous problem and its Fourier transform state, viz b=0
 (cf. Eq.(\ref{mxel})), with the momentum, p, we have that the line of the finite dimension problem
 is parameterized with "initial" values of "q" and "p" i.e. m(-1) and m(0).\\

\section{Mapping onto phase space}

We now define a mapping of Hilbert space operators, e.g. an arbitrary operator, B, onto
the phase space - like lines of DAPG. The mapping is defined by  \cite{revzen1},
\begin{equation}
B\Rightarrow V(j;B)\equiv tr BP_j.
\end{equation}
Here $P_j$ is a line operator within DAPG. (Alternatively we could have cast the mappings within
APG as is clear from the discussion in the previous section.)
The density operator may be expressed in terms of $V(j;\rho)$:
\begin{equation}
\rho=\frac{1}{d}\sum^{d^2}_j\big(tr\rho P_j\big)P_j.
\end{equation}
  $V(j;\rho)$ is quasi distribution \cite{revzen1} in
the phase space like lines and points of DAPG. Thus, for example, the expectation value of
an arbitrary operator $B$ we have,
\begin{equation}
tr\rho B=\frac{1}{d}\sum^{d^2}V(j;\rho)V(j;B).
\end{equation}
The quasi distribution may be reconstructed from the expectation values of the point operator
 $A_{\alpha}$ i.e. MUB state projector's expectation value (obtained, e.g. by measurements),
\begin{equation}
tr \rho A_{\alpha}=\frac{1}{d}\sum^{d^2}V(j;\rho)V(j;A_{\alpha})=
\frac{1}{d}\sum^{d^2}V(j;\rho)\Lambda_{\alpha,j}.
\end{equation}
Thence
\begin{equation}
\frac{1}{d}\sum_{\alpha\in\alpha'\cup M_{\alpha'}}\sum^{d^2}V(j;\rho)\Lambda_{\alpha,j}=V(j;\rho).
\end{equation}

These equations are the finite dimensional Radon transform and its inverse.\\

To summarize we give a comparative list of the central formulae: the expressions for the continuous
phase space vs. the finite plane geometry ones: \\

\noindent a. Phase space mapping,
\begin{eqnarray*}
\rho \Rightarrow&& W_{\rho}(q,p), \\
\rho \Rightarrow&& V_{\rho}(j).
\end{eqnarray*}

\noindent b. MUB state map  ($\alpha$=(m,b))

\begin{eqnarray*}
|x,\theta\rangle\langle\theta,x|\Rightarrow&& W_{|x,\theta\rangle}=\delta(x-Cq-Sp), \\
A_{\alpha} \Rightarrow&& trA_{\alpha}P_j=\Lambda_{\alpha,j}.
\end{eqnarray*}
\noindent c. Radon transform  ($\alpha$=(m,b))
\begin{eqnarray}
{\cal{R}}[W_{\rho}](x,\theta)&=&\int \frac{dqdp}{2\pi}W_{\rho}(q,p)\delta(x-Cq-Sp),\\
{\cal{R}}[V_{\rho}](j)&=&\frac{1}{d}\sum^{d^2}V_{\rho}(j)\Lambda_{\alpha,j}.
\end{eqnarray}
Thus mapping of finite dimensional Hilbert space operators onto phase space-like lines and
points of finite geometry was used to define a finite dimensional phase space Physics and
applied to give the Radon transform and its inversion.

\section{Summary and Concluding Remarks}

Finite geometry stipulates interrelations among lines and points. The stipulations for the
(finite) dual affine plane geometry (DAPG) was shown to conveniently accommodate
association of geometric lines and points with projectors of states of mutual unbiased
bases (MUB). The latter act in a (finite dimensional, d) Hilbert space. This underpinning
of Hilbert space operator with DAPG reveal some novel inter operators relations. Noteworthy
among these are Hilbert space operators, $\hat{P}_j,\;j=1,2,...d^2$, which are underpinned
with DAPG lines, $L_j$: it abides by $\hat{P}_j^2=\hat{I}\;\forall j$, and are mutually
orthogonal, $tr \hat{P}_j\hat{P}_{j'}=d\delta_{j,j'}.$ These allow their utilization for
general mapping of Hilbert space operators onto the phase space like lines and points of
DAPG in close analogy with the mappings within the continuum of Hilbert space operators
onto phase space via the well known Wigner function \cite{moyal,ellinas}. The physics of
phase space involves Weyl-Wigner (W-W) mapping of  Hilbert space operators to functions in
phase space. The W-W image for $\tilde{\rho}(x,\theta)$ is the Radon transform of the
Wigner function, $W_{\rho}(q,p).$ This comes about because the W-W map of the mutual
unbiased state (MUB) involved, $|x,\theta\rangle$, is $\delta(x-Cq-Sp)$ which specifies the
phase space point, q,p, that lie on the line stipulated within the $\delta$ function.
Thence the state reconstruction, gotten via the inversion of the Radon transform requires
the properly normalized sum (integral) of all the lines going through that phase space
point. This involves the handling of singularity in the continuum problem. Finite
dimensional mapping onto the phase space-like lines of the dual affine plane geometry
(DAPG) involves, likewise, mapping of an MUB state, $|m,b\rangle$. This map now relates
DAPG points $\alpha \equiv (m,b)$ and a lines ,j, via $\Lambda_{\alpha,j}$
(Eq.(\ref{del1})) which replaces the continuum $\delta$ function. Mapping and inversion
in the finite dimensional analysis, in close analogy to the Wigner function of the
continuum, is conveniently given in terms of a quasi distribution, $V_{\rho}(j)$ -
the image of the state $\rho$.\\
The approach allows an easy transcription from dual affine plane geometry (DAPG) to affine
plane geometry (APG) and vice versa.

\section*[Appendix A]{Appendix: "Fluctuation Distillation" Formula}

Given,  Eq(\ref{line}),   $\hat{P_j}=\sum_{\alpha \in j}\hat{A_\alpha}-\hat{I}$ and,
  \cite{revzen1}, $\hat{P_j}^2=\hat{I},$ implies

$$\big( \sum_{\alpha \in j}\hat{A_\alpha}-\hat{I}\big) \big(\sum_{\alpha' \in j}\hat{A_\alpha'}
-\hat{I} \big)=\hat{I}.$$ Thus,
$$\sum_{\alpha,\alpha' \in j}\hat{A_\alpha}\hat{A_\alpha'}=2\sum_{\alpha \in j}\hat{A_\alpha}.$$
Recalling that, Eq(\ref{ptop}), $A_{\alpha}^2=A_{\alpha}$ allows
$$\sum_{\alpha \ne\alpha' \in
j}\hat{A_\alpha}\hat{A_\alpha'}=\sum_{\alpha \in j}\hat{A_\alpha}.$$ QED

\end{document}